\documentclass[twocolumn,letter]{jpsj3}
\usepackage{txfonts}
\usepackage{amsmath}
\usepackage{amssymb}
\usepackage{color}
\usepackage{float}
\usepackage{graphicx}

\title{Incommensurate Antiferromagnetic Order in the Fe-substituted Bi-2201 Cuprate in the Heavily Overdoped Regime}

\author{Yota Komiyama$^{1,2}$, Yoichi Ikeda$^3$, Takanori Taniguchi$^3$, Travis J. Williams$^4$, Masaaki Matsuda$^5$, Shinichiro Asai$^6$, Takatsugu Masuda$^6$, Hideki Kuwahara$^1$, Haruhiko Kuroe$^1$, Takayuki Kawamata$^7$, Isao Watanabe$^2$, Masaki Fujita$^3$, and Tadashi Adachi$^1$\thanks{E-mail: t-adachi@sophia.ac.jp}}
\inst{$^1$Department of Engineering and Applied Sciences, Sophia University, Tokyo 102-8554, Japan \\
$^2$Nuclear Structure Research Group, RIKEN Nishina Center, Wako 351-0198, Japan \\
$^3$Institute for Materials Research, Tohoku University, Sendai 980-8577, Japan \\
$^4$ISIS Pulsed Neutron and Muon Source, STFC Rutherford Appleton Laboratory, Harwell Campus, Didcot, Oxfordshire OX11 0QX, United Kingdom \\
$^5$Neutron Scattering Division, Oak Ridge National Laboratory, Tennessee 37831, U.S.A. \\
$^6$Institute for Solid State Physics, The University of Tokyo, Kashiwa 277-8581, Japan \\
$^7$Department of Natural Sciences, Tokyo Denki University, Tokyo 120-8551, Japan\\ }

\abst{Elastic neutron scattering experiments showed incommensurate antiferromagnetic peaks in 5\% Fe-substituted Bi-2201 cuprate in the non-superconducting heavily overdoped regime. The incommensurability $\delta \sim 0.21$ is comparable to that observed in Fe-substituted Bi-2201 in the overdoped regime. [Hiraka $et$ $al$., Phys. Rev. B ${\bf 81}$, 144501 (2010).] The magnetic correlation length is comparable between the overdoped and non-superconducting heavily overdoped regimes. It is plausible that incommensurate antiferromagnetic order is induced and stabilized by Fe in the heavily overdoped regime, which suggests a robust antiferromagnetic correlation beyond the superconducting dome in the phase diagram.}

\kword{Bi-2201, incommensurate antiferromagnetic order, heavily overdoped regime, magnetic impurity}

\begin{document}
\maketitle

The mechanism of high-$T_{\rm c}$ cuprate superconductivity has been a long-standing problem. A prevailing idea is that antiferromagnetic (AFM) fluctuations play a role in forming the superconducting electron pairs in such cuprates \cite{RJBirgeneau}. Since the early stages of the discovery of high-$T_{\rm c}$ cuprates, incommensurate (IC) AFM fluctuations have been observed in the hole-doped cuprate La$_{2-x}$Sr$_{x}$CuO$_{4}$ (LSCO) \cite{SWCheong,KYamada}, YBa$_{2}$Cu$_{3}$O$_{6+d}$ (YBCO) \cite{JMTranquada1,PDai}, and Bi$_{2}$Sr$_{2}$CaCu$_{2}$O$_{8+d}$ \cite{HFFong,HHe} in the underdoped and optimally doped regimes. In LSCO \cite{KYamada,MMatsuda,MFujita1} and YBa$_{2}$Cu$_{3}$O$_{7-d}$ \cite{PDai}, it was reported that the incommensurability $\delta$ is proportional to the hole concentration per Cu $p$ and the superconducting transition temperature $T_{\rm c}$ in the underdoped regime, which suggests an intimate relation between IC-AFM fluctuations and superconductivity. A theoretical proposal was that the formation of Cooper pairs is meditated by the IC-AFM fluctuations \cite{VJEmery}.

In the overdoped regime in which superconductivity is gradually suppressed by doping, inelastic neutron-scattering experiments with LSCO revealed that the IC-AFM fluctuations were weakened with overdoping and disappeared in the heavily overdoped (HOD) regime where the superconductivity was suppressed, which suggested that the suppression of superconductivity in the HOD regime was attributed to the disappearance of the IC-AFM fluctuations \cite{SWakimoto1}. From muon-spin-relaxation ($\mu$SR) measurements of La$_{2-x}$Sr$_{x}$Cu$_{1-y}$Zn$_{y}$O$_{4}$ \cite{Risdiana} and Bi$_{1.74}$Pb$_{0.38}$Sr$_{1.88}$Cu$_{1-y}$Zn$_{y}$O$_{6+d}$ \cite{TAdachi1} with non-magnetic Zn-substitution, the relaxation rate of muon spins was observed to decrease with increasing $p$ in the overdoped regime, which suggested that the suppression of superconductivity in the overdoped regime was due to weakening of the AFM fluctuations. On the other hand, resonant inelastic X-ray scattering (RIXS) revealed the existence of high-energy AFM fluctuations in HOD LSCO \cite{MPMDean}. Therefore, the relationship between the suppression of superconductivity and AFM fluctuations has not yet been clarified.

Ferromagnetism is another candidate for the origin of the suppression of superconductivity in the HOD regime for cuprates. Theoretical studies \cite{AKopp,CJJia,TAMaier} have proposed ferromagnetic order/fluctuations in the non-superconducting HOD regime for Tl$_{2}$Ba$_{2}$CuO$_{6+d}$ and LSCO. It was also proposed that the spin susceptibility increased at $q$ $\sim$ (0, 0) with overdoping \cite{STeranishi} and that local magnetic moments due to Hund's coupling were enhanced in the HOD regime \cite{HWatanabe}, which suggests the enhancement of ferromagnetic fluctuations. Measurements of transport, magnetization, $\mu$SR and RIXS suggested ferromagnetic order/fluctuations in the HOD regime of LSCO \cite{JESonier}, Bi-2201 \cite{KKurashima,YYPeng,HRaffy}, Tl$_{2}$Ba$_{2}$CuO$_{6+d}$ \cite{TAdachi2} and electron-doped La$_{2-x}$Ce$_{x}$CuO$_{4}$ \cite{TSakar}. Anomalous rotational anisotropy of second-harmonic generation in overdoped Bi-2212 may be related to ferromagnetism \cite{SJung}. However, the relationship between AFM fluctuations, ferromagnetic fluctuations and superconductivity is yet unknown.

Impurity substitution for Cu has been a powerful tool to investigate the magnetism in the cuprates. $\mu$SR measurements of Zn-substituted La$_{2-x}$Sr$_{x}$Cu$_{1-y}$Zn$_{y}$O$_{4}$ \cite{Risdiana,IWatanabe,TAdachi3,TAdachi4} and Bi$_{1.74}$Pb$_{0.38}$Sr$_{1.88}$Cu$_{1-y}$Zn$_{y}$O$_{6+d}$ \cite{TAdachi1} revealed that the relaxation rate of muon spins was enhanced by Zn in a wide doping range where superconductivity appeared, which suggested that the stripe fluctuations were pinned and stabilized around Zn. For the magnetic-impurity substitution, $\mu$SR measurements of magnetic-Fe-substituted La$_{2-x}$Sr$_{x}$Cu$_{1-y}$Fe$_{y}$O$_{4}$ revealed that the stripe fluctuations were enhanced around $p$ = 1/8 more than in the case of Zn substitution \cite{KMSuzuki}. On the other hand, $\delta$ in overdoped La$_{2-x}$Sr$_{x}$Cu$_{1-y}$Fe$_{y}$O$_{4}$ obtained from neutron scattering was comparable to that calculated using the Lindhard susceptibility based on the Fermi surface, which suggested that the IC-AFM order in the overdoped regime originated from the Ruderman-Kittel-Kasuya-Yosida (RKKY) interaction \cite{RHHe}. In the overdoped regime of Fe-substituted Bi-2201, IC-AFM peaks with $\delta$ $\sim$ 0.20 were observed from elastic neutron scattering experiments \cite{HHiraka}. The peak intensity was weakened by the application of a magnetic field and negative magnetoresistance was observed; therefore, it was proposed that the Kondo effect and RKKY interaction were the cause of the IC-AFM correlation \cite{SWakimoto2}. However, $\delta$ for overdoped Bi-2201 is close to the value expected from the proportional relationship between $\delta$ and $p$; therefore, stripe-related magnetism cannot be excluded. Our previous magnetic-susceptibility and electrical-resistivity measurements of the Fe-substituted Bi-2201 cuprate Bi$_{1.74}$Pb$_{0.38}$Sr$_{1.88}$Cu$_{1-y}$Fe$_{y}$O$_{6+d}$ ($y$ = 0 -- 0.09) in the HOD regime \cite{YKomiyama} revealed that the onset temperature for a spin-glass state increased linearly with Fe content, that the effective Bohr magneton calculated from the Curie constant was larger than that for a Fe$^{3+}$ spin, and that the dimensionality of ferromagnetic fluctuations changed from two to three dimensions, which suggested that the ferromagnetic fluctuations were stabilized by Fe substitution. However, it is unclear whether the IC-AFM order is formed in the HOD regime of Fe-substituted Bi-2201. In this paper, we investigate the details of the magnetism in the HOD regime of 5\% Fe-substituted Bi-2201 by neutron scattering measurments.


\begin{figure}[tbp]
\begin{center}
\includegraphics[width=1.0\linewidth,clip]{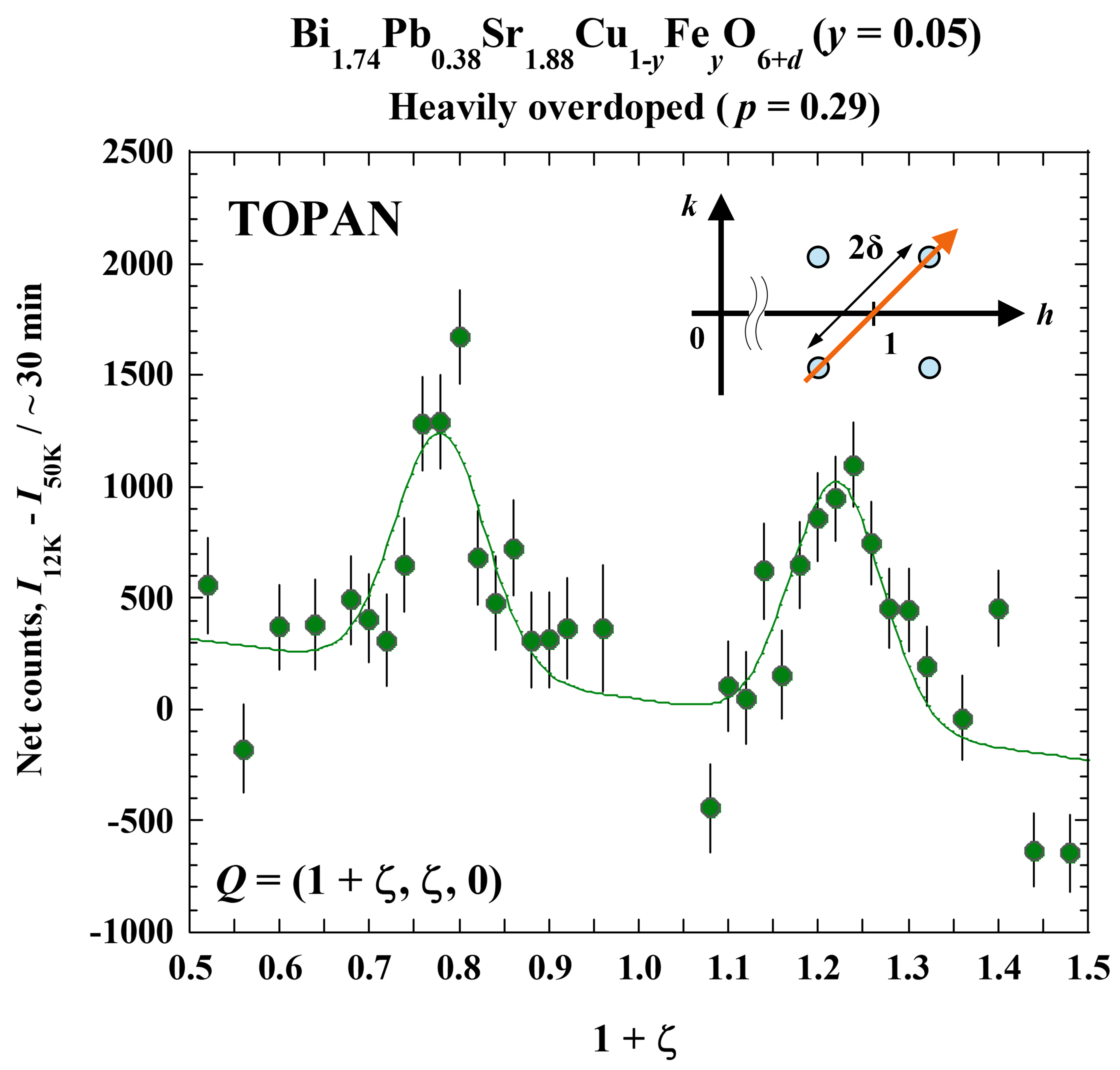}
\caption{(Color online) Difference profile obtained by subtracting the data at 50 K from those at 12 K in HOD Bi$_{1.74}$Pb$_{0.38}$Sr$_{1.88}$Cu$_{1-y}$Fe$_{y}$O$_{6+d}$ with $y$ = 0.05 ($p$ = 0.29) using TOPAN. Inset shows the scan trajectory in the reciprocal ($h$, $k$) plane of the orthorhombic notation. Solid line is a fitting result using Eq. (1).}
\label{fig:f1}
\end{center}
\end{figure}

Single crystals of Bi$_{1.74}$Pb$_{0.38}$Sr$_{1.88}$Cu$_{1-y}$Fe$_{y}$O$_{6+d}$ ($y$ = 0.05) were grown by the floating-zone method \cite{KKurashima,YKomiyama}. X-ray diffraction analysis of the obtained crystals indicated a single phase. Two crystal rods were annealed in flowing oxygen at 1 atm; one for 50 h and the other for 120 h. The $p$ values estimated from the Seebeck coefficient at 290 K \cite{SDObertelli} were 0.29(0) and 0.31(0) for the 50 h- and 120 h- annealed crystal rods, respectively, which indicates that the samples resided in the HOD regime. Crystal rods with a total weight of $\sim$ 2 g were used for neutron-scattering experiments. The quality and orientation of the crystals were determined using the triple-axis AKANE spectrometer at the Japan Research Reactor No. 3, Tokai, Japan (JRR-3). Elastic neutron scattering measurements were conducted using the triple-axis, TOPAN and HER spectrometers at JRR-3 for the same 50 h-annealed crystals and HB-3 installed at the High Flux Isotope Reactor (HFIR) at the Oak Ridge National Laboratory, USA, for 120 h-annealed crystals. The typical energy resolutions of the instruments were $\sim$ 1.5 meV for TOPAN and HB-3, and $\sim$ 0.1 meV for HER.


\begin{figure}[tbp]
\begin{center}
\includegraphics[width=1.0\linewidth,clip]{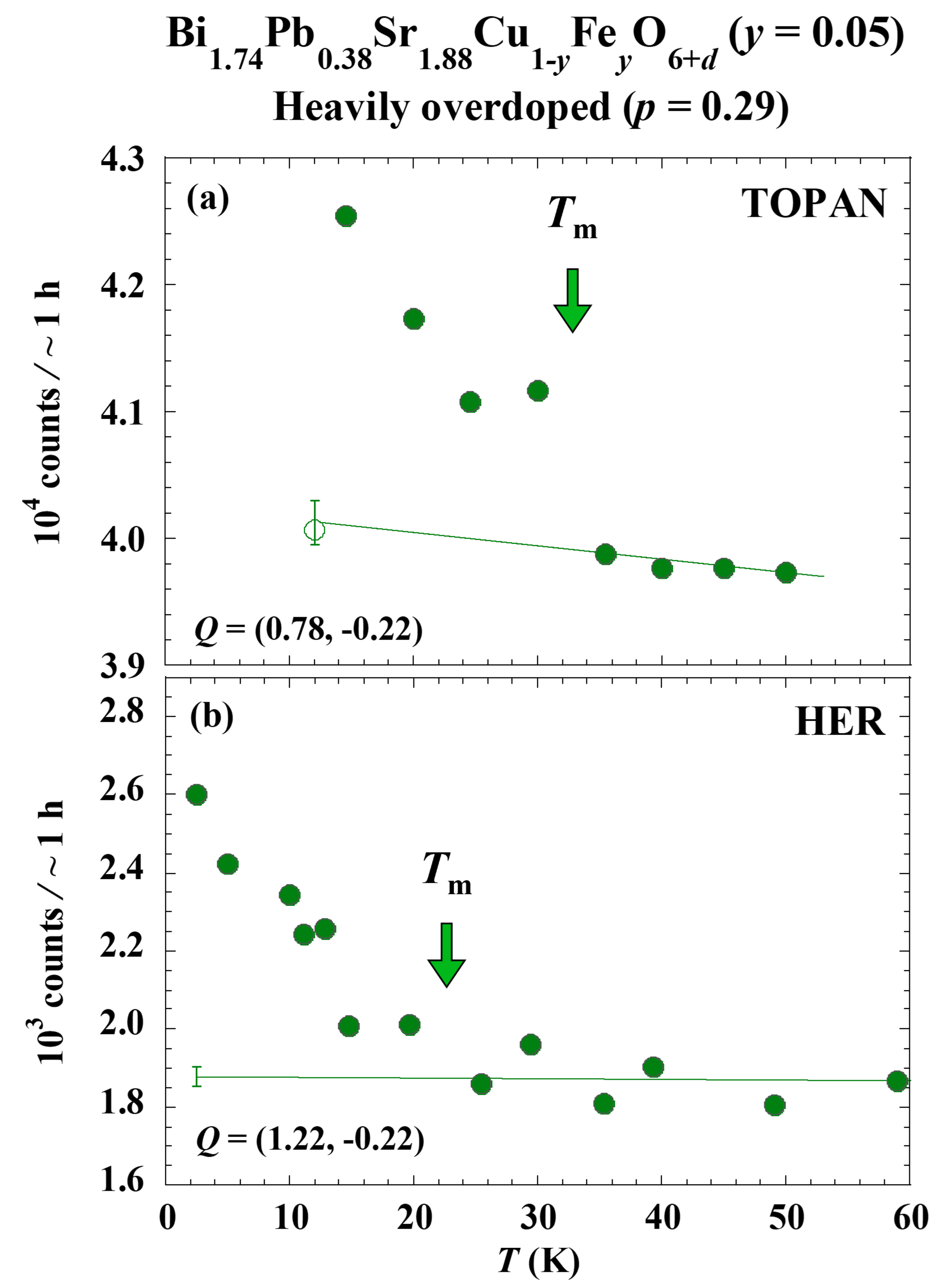}
\caption{(Color online) Temperature dependence of net counts at the IC peak position for HOD Bi$_{1.74}$Pb$_{0.38}$Sr$_{1.88}$Cu$_{1-y}$Fe$_{y}$O$_{6+d}$ with $y$ = 0.05 ($p$ = 0.29) for (a) TOPAN (energy resolution: $\sim$ 1.5 meV) and (b) HER ($\sim$ 0.1 meV). Solid lines represent the background, derived from the temperature-dependent intensity changes at the ($h,k$) position unaffected by the IC peaks, extrapolated from the data at the highest temperature. Error bars at the end of background lines represent errors in the intensity change. The open circle in (a) represents the background value at $Q=$ (0.78,-0.22) in Fig. 1 added to the 50 K data. Arrows indicate the magnetic transition temperature $T_{\rm m}$ at which the net counts begin to deviate from the background. Details are described in the text.}
\label{fig:f2}
\end{center}
\end{figure}

Figure 1 shows a difference elastic neutron-scattering profile in HOD Bi$_{1.74}$Pb$_{0.38}$Sr$_{1.88}$Cu$_{1-y}$Fe$_{y}$O$_{6+d}$ with $y$ = 0.05 ($p$ = 0.29) obtained by subtracting the data at 50 K from those at the lowest-temperature of 12 K using TOPAN, and by scanning in the reciprocal ($h$, $k$)$_{\rm ortho}$ plane of the orthorhombic notation. Incommensurate Bragg reflections were observed around (1 $\pm$ $\delta$, 0 $\pm$ $\delta$)$_{\rm ortho}$, which indicates the formation of IC-AFM order at low temperatures. The IC-AFM peaks were also observed using HER with better energy resolution and at HB-3 using samples with $p$ = 0.31. The obtained profile was fitted using the following Gaussian function,
\begin{equation}
I(1+\zeta) = h_{1} \times \exp\left[-\frac{(\zeta - \delta)^2}{2\sigma^2}\right] + h_{2} \times \exp\left[-\frac{(\zeta + \delta)^2}{2\sigma^2}\right] + a \times \zeta + b,
\end{equation}
where $h_1$ and $h_2$ are the peak heights, $\delta$ is the incommensurability, $\sigma$ is the standard deviation related to the full width at half maximum (FWHM) using the relation FWHM = 2$\sqrt{2\rm{ln}2}\sigma$, and a $\times$ $\zeta$+b is the linear background.

Figure 2 shows the temperature dependence of the net count at the IC-AFM position for HOD Bi$_{1.74}$Pb$_{0.38}$Sr$_{1.88}$Cu$_{1-y}$Fe$_{y}$O$_{6+d}$ ($y$ = 0.05) obtained with TOPAN and HER. Solid lines represent the background derived from the temperature-dependent intensity outside the peak position. This estimation is reasonable because the background value at $Q=$ (0.78, -0.22) in Fig. 1 added to the 50 K data falls within the range of the error bars. It is found that the net count gradually increases with decreasing temperature at low temperatures and shows no sign of saturation. Similar behavior has also been observed in Zn-substituted La$_{1.85}$Sr$_{0.15}$Cu$_{1-y}$Zn$_{y}$O$_{4}$ \cite{HKimura} and Fe-substituted La$_{2-x}$Sr$_{x}$Cu$_{1-y}$Fe$_{y}$O$_{4}$ \cite{RHHe}, which suggests the development of glassy magnetism in HOD Bi$_{1.74}$Pb$_{0.38}$Sr$_{1.88}$Cu$_{1-y}$Fe$_{y}$O$_{6+d}$ ($y$ = 0.05). 
The magnetic transition temperature $T_{\rm m}$ is defined as $T_{\rm m}$ = ($T_{\rm m1}$ + $T_{\rm m2}$) / 2 where $T_{\rm m1}$ is the temperature at which the net counts ($N_{\rm 1}$ - $N_{\rm 2}$) exceeds the statistical error $\sqrt{N_1 + N_2}$ assuming the Poisson statistics, and $T_{\rm m2}$ is the highest temperature at which the signal remains within this error. Here, $N_1$ and $N_2$ denote the net counts and the background intensity, respectively. The error of $T_{\rm m}$ is considered as $\Delta T_{\rm m} = T_{\rm m1} - T_{\rm m}$. With this definition, $T_{\rm m}$ was determined to be 32.8 $\pm$ 2.8 K and 22.6 $\pm$ 2.9 K using TOPAN and HER, respectively.

\begin{figure}[tbp]
\begin{center}
\centering
\includegraphics[width=1.0\linewidth]{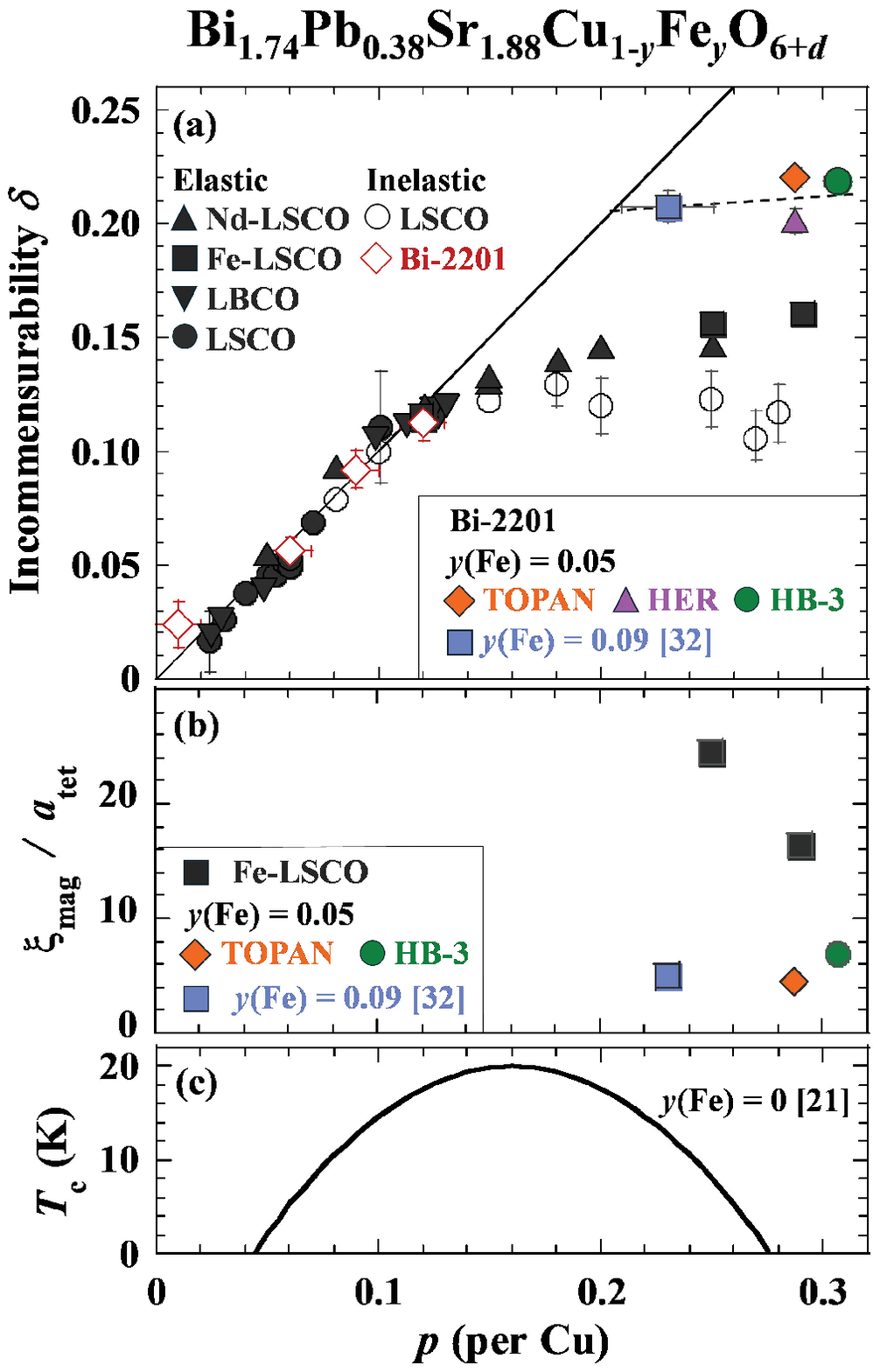}
\caption{(Color online) (a) Hole-concentration dependence of incommensurability $\delta$ in HOD Bi$_{1.74}$Pb$_{0.38}$Sr$_{1.88}$Cu$_{1-y}$Fe$_{y}$O$_{6+d}$ ($y$ = 0.05) obtained with TOPAN, HER, and HB-3, together with overdoped Bi$_{1.75}$Pb$_{0.35}$Sr$_{1.90}$Cu$_{1-y}$Fe$_{y}$O$_{6+d}$ ($y$ = 0.09) \cite{HHiraka} analyzed using the same procedure as in the present study, La$_{2-x}$Ba$_{x}$CuO$_{4}$ \cite{MFujita2,SRDunsiger1,SRDunsiger2}, La$_{2-x}$Sr$_{x}$CuO$_{4}$ \cite{MMatsuda,HKimura2,SWakimoto3,SWakimoto4}, La$_{1.6-x}$Nd$_{0.4}$Sr$_{x}$CuO$_{4}$ \cite{JMTranquada2,JMTranquada3,NIchikawa,SWakimoto5} and La$_{2-x}$Sr$_{x}$Cu$_{1-y}$Fe$_{y}$O$_{4}$ ($y$ = 0.01)\cite{RHHe}, obtained in the elastic scan, and Bi$_{2+x}$Sr$_{2-x}$CuO$_{6+d}$ \cite{MEnoki} and La$_{2-x}$Sr$_{x}$CuO$_{4}$ \cite{KYamada,SWakimoto1,CHLee} in the inelastic scan. Closed (open) symbols denote the data obtained in the elastic (inelastic) scan. The solid line represents the linear relationship $\delta$ = $p$. The dashed line indicates the trend for the Fe-substituted Bi-2201 data. (b) Hole-concentration dependence of the magnetic correlation length $\xi_{\rm mag}$ divided by the tetragonal a-axis lattice constant $a_{\rm tet}$ in Bi$_{1.74}$Pb$_{0.38}$Sr$_{1.88}$Cu$_{1-y}$Fe$_{y}$O$_{6+d}$ ($y$ = 0.05), overdoped Bi$_{1.75}$Pb$_{0.35}$Sr$_{1.90}$Cu$_{1-y}$Fe$_{y}$O$_{6+d}$ ($y$ = 0.09) obtained by analysis using Eq. (1),\cite{HHiraka} and La$_{2-x}$Sr$_{x}$Cu$_{1-y}$Fe$_{y}$O$_{4}$ ($y$ = 0.01)\cite{RHHe}. (c) Hole-concentration dependence of $T_{\rm c}$ in Fe-free Bi$_{1.74}$Pb$_{0.38}$Sr$_{1.88}$Cu$_{1-y}$Fe$_{y}$O$_{6+d}$ ($y$ = 0) \cite{KKurashima}. }
\label{fig:f3}
\end{center}
\end{figure}

Figure 3(a) shows the hole-concentration dependence of $\delta$ in HOD Bi$_{1.74}$Pb$_{0.38}$Sr$_{1.88}$Cu$_{1-y}$Fe$_{y}$O$_{6+d}$ ($y$ = 0.05) obtained with TOPAN, HER, and HB-3, together with overdoped Bi$_{1.75}$Pb$_{0.35}$Sr$_{1.90}$Cu$_{1-y}$Fe$_{y}$O$_{6+d}$ ($y$ = 0.09) obtained by analysis using Eq. (1) \cite{HHiraka}, La$_{2-x}$Ba$_{x}$CuO$_{4}$ \cite{MFujita2,SRDunsiger1,SRDunsiger2}, La$_{2-x}$Sr$_{x}$CuO$_{4}$ \cite{MMatsuda,HKimura2,SWakimoto3,SWakimoto4}, La$_{1.6-x}$Nd$_{0.4}$Sr$_{x}$CuO$_{4}$ \cite{JMTranquada2,JMTranquada3,NIchikawa,SWakimoto5} and La$_{2-x}$Sr$_{x}$Cu$_{1-y}$Fe$_{y}$O$_{4}$ ($y$ = 0.01)\cite{RHHe} obtained in the elastic scans, and Bi$_{2+x}$Sr$_{2-x}$CuO$_{6+d}$ \cite{MEnoki} and La$_{2-x}$Sr$_{x}$CuO$_{4}$ \cite{KYamada,SWakimoto1,CHLee} obtained in the inelastic scans. Below $p$ $\sim$ 1/8, $\delta$ tends to increase linearly with $p$ in Bi-2201 and the La-based cuprates. Above $p$ $\sim$ 1/8, while $\delta$ for LSCO obtained from inelastic scattering tends to saturate, $\delta$ for La$_{1.6-x}$Nd$_{0.4}$Sr$_{x}$CuO$_{4}$ and Fe-substituted La$_{2-x}$Sr$_{x}$Cu$_{1-y}$Fe$_{y}$O$_{4}$ obtained from elastic scattering increases gradually. The $\delta$ values for HOD Bi$_{1.74}$Pb$_{0.38}$Sr$_{1.88}$Cu$_{1-y}$Fe$_{y}$O$_{6+d}$ ($y$ = 0.05) are larger than those for the La-based cuprates \cite{RHHe,JMTranquada2,JMTranquada3,NIchikawa,SWakimoto5} and are comparable to that for overdoped Bi$_{1.74}$Pb$_{0.38}$Sr$_{1.88}$Cu$_{1-y}$Fe$_{y}$O$_{6+d}$ ($y$ = 0.09) \cite{HHiraka}. These results suggest the formation of an identical IC-AFM order between the HOD and overdoped regimes of Bi-2201. Figure 3(b) shows the magnetic correlation length $\xi_{\rm mag}$ obtained using TOPAN and HB-3 with the same instrumental energy resolution, and calculated using the formula $\xi_{\rm mag}$ = $1/(\sqrt{2\ln2}\sigma)$, which is comparable between the overdoped and HOD samples.

IC-AFM peaks were observed in 5\% Fe-substituted Bi-2201 in the HOD regime. While a magnetic order was not detected in impurity-free Bi-2201 in the HOD regime \cite{TAdachi1}, the emergence of an IC-AFM order upon Fe substitution suggests that AFM fluctuations exist in Fe-free Bi-2201 in the HOD regime. $T_{\rm m}$ is dependent on the energy resolution of the spectrometers. i.e., $T_{\rm m}$ is $32.8 \pm 2.8$ K using TOPAN with an energy resolution of $\sim$ 1.5 meV, while $T_{\rm m}$ is $22.6 \pm 2.9$ K using HER with an energy resolution of $\sim$ 0.1 meV. This resolution dependent behavior of $T_{\rm m}$ reflects the time scale of spin fluctuations probed by each spectrometer. A similar phenomenon has been reported in the triangular lattice magnet NiGa$_{2}$S$_{4}$ \cite{YNambu}. Moreover, our former $\mu$SR measurements that probed lower-energy spin fluctuations than neutron scattering of 5\% Fe-substituted Bi$_{1.74}$Pb$_{0.38}$Sr$_{1.88}$Cu$_{1-y}$Fe$_{y}$O$_{6+d}$ ($y$ = 0.05) \cite{TAdachi5} revealed a magnetic transition temperature of $\sim$ 6 K. These results suggest quasistatic IC-AFM in nature and the formation of glassy magnetism as observed 1\% Fe-substituted La$_{2-x}$Sr$_{x}$Cu$_{1-y}$Fe$_{y}$O$_{4}$ ($y$ = 0.01) \cite{RHHe}. This is also observable from the temperature dependence of the net count shown in Fig. 2, where the net count increases gradually with decreasing temperature below $T_{\rm m}$, which is different from the order-parameter-like behavior. The $\delta$ is almost identical to and $\xi_{\rm mag}$ is comparable to those for overdoped Bi-2201 with 9\% Fe substitution, which suggests that the IC-AFM order has the same origin for the overdoped and HOD regimes. As shown in Fig. 3(a), $\delta$ is in proportion to $p$ for 0 $<$ $p$ $\lesssim$ 1/8 in the La-based cuprates, which is related to the stripe-like magnetism \cite{MMatsuda,MFujita2,SRDunsiger1,SRDunsiger2,HKimura2,SWakimoto3,SWakimoto4}. Above $p$ $\sim$ 1/8, $\delta$ deviates from the relation $\delta$ = $p$ and increases gradually toward the non-superconducting HOD regime \cite{RHHe,JMTranquada2,JMTranquada3,NIchikawa,SWakimoto5}. It was suggested that the spin-density-wave (SDW) order due to RKKY interaction in 1\% Fe-substituted La$_{2-x}$Sr$_{x}$Cu$_{1-y}$Fe$_{y}$O$_{4}$ \cite{RHHe} was formed in the overdoped regime. Moreover, hysteresis in the temperature dependence of the magnetic susceptibility and $\mu$SR results suggest the novel coexistence of spin glass and SDW in 1\% Fe-substituted La$_{2-x}$Sr$_{x}$Cu$_{1-y}$Fe$_{y}$O$_{4}$ in the overdoped regime \cite{KMSuzuki2}. For Fe-substituted Bi-2201 in the overdoped and HOD regimes, $\delta$ is larger than that for the La-based cuprates. The tendency for $\delta$ in Bi-2201 to saturate in the overdoped and HOD regimes is similar to that in the La-based cuprates; therefore it is possible that the novel coexistence of spin glass and SDW induced by the RKKY interaction is also realized in overdoped and HOD Bi-2201.

Compared with 1\% Fe-substituted La$_{2-x}$Sr$_{x}$Cu$_{1-y}$Fe$_{y}$O$_{4}$ in the HOD regime \cite{RHHe}, the IC-AFM peaks are broad and $\xi_{\rm mag}$ is short in 5\% Fe-substituted Bi$_{1.74}$Pb$_{0.38}$Sr$_{1.88}$Cu$_{1-y}$Fe$_{y}$O$_{6+d}$ ($y$ = 0.05), as shown in Figure 3(b). A similar tendency in $\xi_{\rm mag}$ between Fe-free LSCO and Bi-2201 is also observed in the underdoped regime \cite{KYamada,SWakimoto1,MEnoki,CHLee}. In the underdoped regime, elastic IC-AFM peaks were observed for LSCO, while IC-AFM peaks were observed only for inelastic scattering in Bi-2201 \cite{MEnoki}. $T_{\rm m}$ for the present 5\% Fe-substituted Bi-2201 was observed to be lower than that for LSCO \cite{RHHe}. These results suggest that the IC-AFM correlation in Bi-2201 is weaker than that in LSCO, which could result from the crystal structure of Bi-2201 being more two-dimensional than that of LSCO.

Elastic neutron scattering experiments revealed the formation of AFM order through Fe substitution, and the magnetic correlation length was comparable to the average distance between Fe atoms. In addition, magnetic-susceptibility measurements revealed that the Fe substitution also leads to the formation of a spin glass, and the spin-glass transition temperature increases linearly with Fe content\cite{YKomiyama}. These results are explained in terms of the AFM spin cluster, i.e., the AFM order is formed around Fe in a cluster-like manner, and the magnetic moments generated by each cluster give rise to a spin glass. This has also been suggested in overdoped Bi-2201 with Fe substitution \cite{HHiraka}. Accordingly, it is suggested that IC-AFM fluctuations are stabilized around Fe, which leads to the formation of AFM clusters.

Ferromagnetic fluctuations have been proposed in HOD Bi-2201 \cite{KKurashima,YYPeng,HRaffy}. Moreover, magnetic susceptibility results suggested the enhancement of ferromagnetic fluctuations by Fe substitution \cite{YKomiyama}. The IC-AFM fluctuations are robust in Fe-free Bi-2201 in the non-superconducting HOD regime; therefore, the mechanism leading to the suppression of superconductivity in the HOD regime for Bi-2201 may be related to ferromagnetic fluctuations. To clarify how the IC-AFM order and ferromagnetic fluctuations coexist in the HOD regime remains a future task.

Elastic neutron-scattering experiment revealed that IC-AFM order was induced by 5\% Fe substitution in HOD Bi-2201 Bi$_{1.74}$Pb$_{0.38}$Sr$_{1.88}$Cu$_{1-y}$Fe$_{y}$O$_{6+d}$ ($y=0.05$). Values of $T_{\rm m}$ were dependent on the energy resolution of spectrometers and the intensity of the IC-AFM peaks tended not to saturate at low temperatures, which suggests the development of glassy magnetism. The value of $\delta$ for HOD Bi-2201 was comparable to that for overdoped Bi-2201, which suggests the same origin of magnetism in both. It is suggested that IC-AFM fluctuations due to itinerant electrons driven by the RKKY interaction are stabilized around Fe, which leads to the formation of AFM clusters. The IC-AFM order is induced by Fe substitution; therefore, IC-AFM fluctuations are presumed to exist in Fe-free Bi-2201 in the HOD regime. Both AFM and ferromagnetic fluctuations would thus coexist in HOD Bi-2201. This raises the possibility that the suppression of superconductivity in the HOD regime may be primarily related to the presence of ferromagnetic fluctuations. Further studies are mandatory to investigate the origin of the IC-AFM order and the relation to ferromagnetic fluctuations in the HOD regime.

\begin{acknowledgment}
The authors would like to thank H. Hiraka for providing the neutron-scattering data for Fe-substituted Bi-2201 in the overdoped regime. We also thank Mr. M. Ohkawara for his kind support with the neutron experiments at JRR-3. Y. K. was supported by the RIKEN Junior Research Associate Program. This research used resources at the High Flux Isotope Reactor, DOE Office of Science User Facility operated by the Oak Ridge National Laboratory. The beam time was allocated to HB-3 on proposal number IPTS-25010. This work was partly supported by the US Japan Collaborative Program on Neutron Scattering. The neutron scattering experiments were performed under the Joint-Use Research Program for Neutron Scattering, Institute for Solid State Physics (ISSP), University of Tokyo, and the Japan Research Reactor JRR-3 (TOPAN on IRT 22402, AKANE on IRT on 24409 and C1-1 on No. 22548). We gratefully acknowledge support from the Center of Neutron Science for Advanced Materials, Institute for Materials Research, Tohoku University (GIMRT: 202112-CNKXX-0004).

\end{acknowledgment}

\end{document}